\documentclass[journal]{IEEEtran}

\ifCLASSINFOpdf
\else
   \usepackage[dvips]{graphicx}
\fi
\usepackage{url}
\usepackage{graphicx}
\usepackage{amssymb}
\usepackage{amsmath}
\usepackage{subfigure}
\usepackage{float}
\usepackage{setspace}

\begin{document}

\title{Reconstruction Condition of Quantized Signals in Unlimited Sampling Framework}

\author{Yan He, Jifang Qiu, Chang Liu, Yue Liu, and  Jian Wu
\thanks{This work was partly supported by the National Natural Science Foundation of China (Grant No. 61935003 and No. 61875020) and the National Key Research and Development Program of China (Grant No. 2019YFB1803601).}
\thanks{The authors are with the School of Electronic Engineering, Beijing University of Posts and Telecommunications, Beijing 100876, P.R. China (email: jifangqiu@bupt.edu.cn).}}

\markboth{}
{Shell \MakeLowercase{\textit{et al.}}: Bare Demo of IEEEtran.cls for IEEE Journals}
\maketitle

\begin{abstract}
The latest theoretical advances in the field of unlimited sampling framework (USF) show the potential to avoid clipping problems of analog-to-digital converters (ADC). To date, most of the related works have focused on real-valued modulo samples, but little has been reported about the impact of quantization. In this paper, we study more practical USF system where modulo samples are quantized to a finite number of bits. In particular, we present a new requirement about the lower bound of sampling rate to ensure exact recovery of original signals from quantized modulo samples. The minimum sampling rate is jointly determined by signal bandwidth and quantization bits. Numerical results show that in the presence of quantization noise, signals with different waveforms and bandwidths are recovered perfectly at the new minimum sampling rate while the recovery fails at minimum sampling rate before modification, which also verifies the correctness of the theory. The trade-offs of sampling rates, quantization bits and computational complexity of recovery algorithm are also given for practitioners to weigh.
\end{abstract}

\begin{IEEEkeywords}
analog-to-digital converter (ADC), high dynamic range, modulo ADC, unlimited sampling, quantization, sampling theory.
\end{IEEEkeywords}

\IEEEpeerreviewmaketitle

\section{Introduction}

\IEEEPARstart{D}{UE} to constraints on the hardware, the so-called clipping or saturation problem exists in data acquisition systems such as analog-to-digital converters (ADC). The saturation errors can be unbounded since large amplitudes that are out of the dynamic range (DR) of ADC are not representable and typically assigned to the values corresponding to the upper or lower limits of quantizers, as shown in Fig. 1(a). Such a problem can be avoided by enhancing the DR of ADC with improved hardware which can be expensive. Unlimited Sampling Framework (USF) gave a novel approach based on the fact that ADC can reset and produce modulo samples if a signal exceeds ADC saturation voltage $\lambda$ [1].

The concept of \emph{modulo} ADC was introduced to describe an ADC that can apply a modulo operation on the input signal [6]. As illustrated in Fig. 1, the input DR of \emph{modulo} ADC is infinite due to the fact that although the modulo mapping distorts signal waveforms, it captures all the variations of input signals which then can be recovered by unwrapping the modulo-folding (Fig. 1(a)), whereas the output signal of conventional ADC saturates to $\lambda$ and this results in clipping (Fig. 1(b)).

\begin{figure}
		\centering
		\subfigure[]{\includegraphics[height=1.9cm,width=4.23cm]{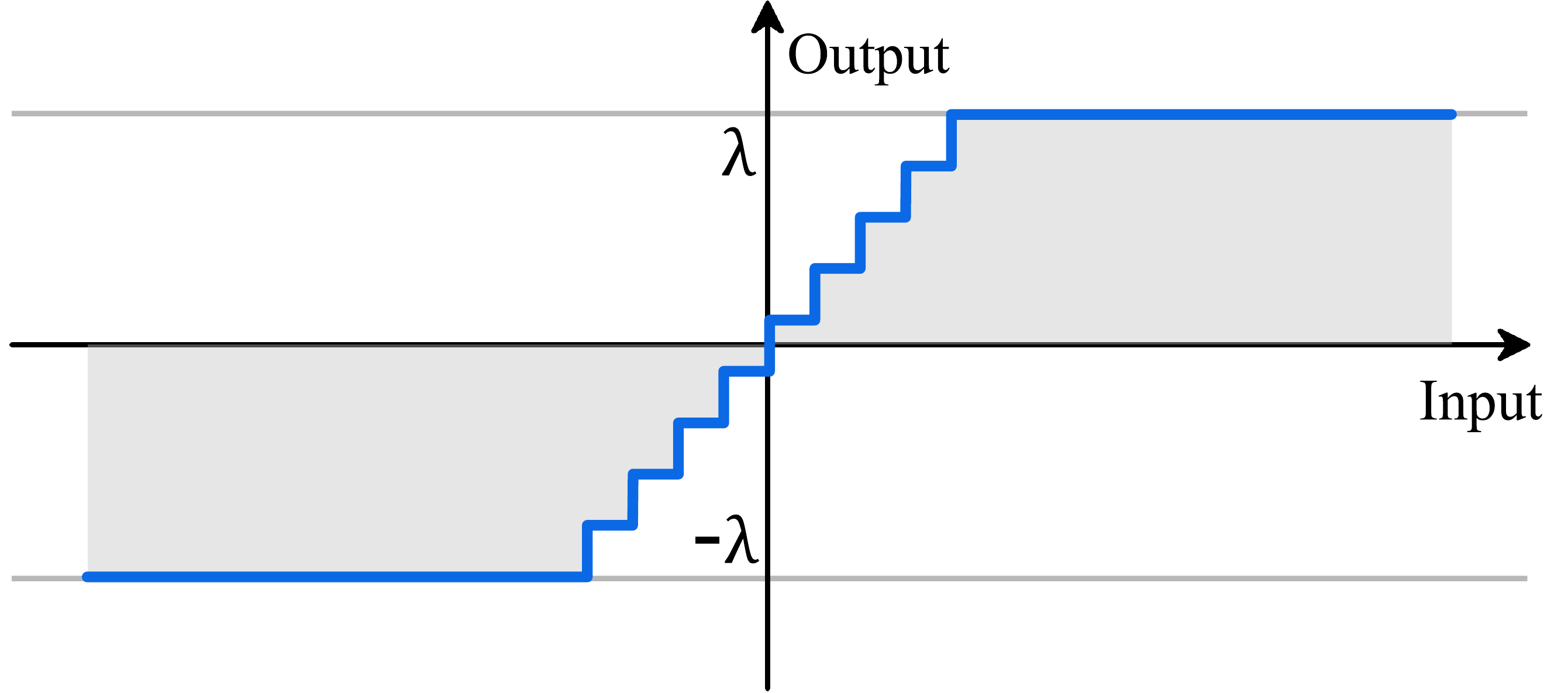}}
\,
		\subfigure[]{\includegraphics[height=1.9cm,width=4.23cm]{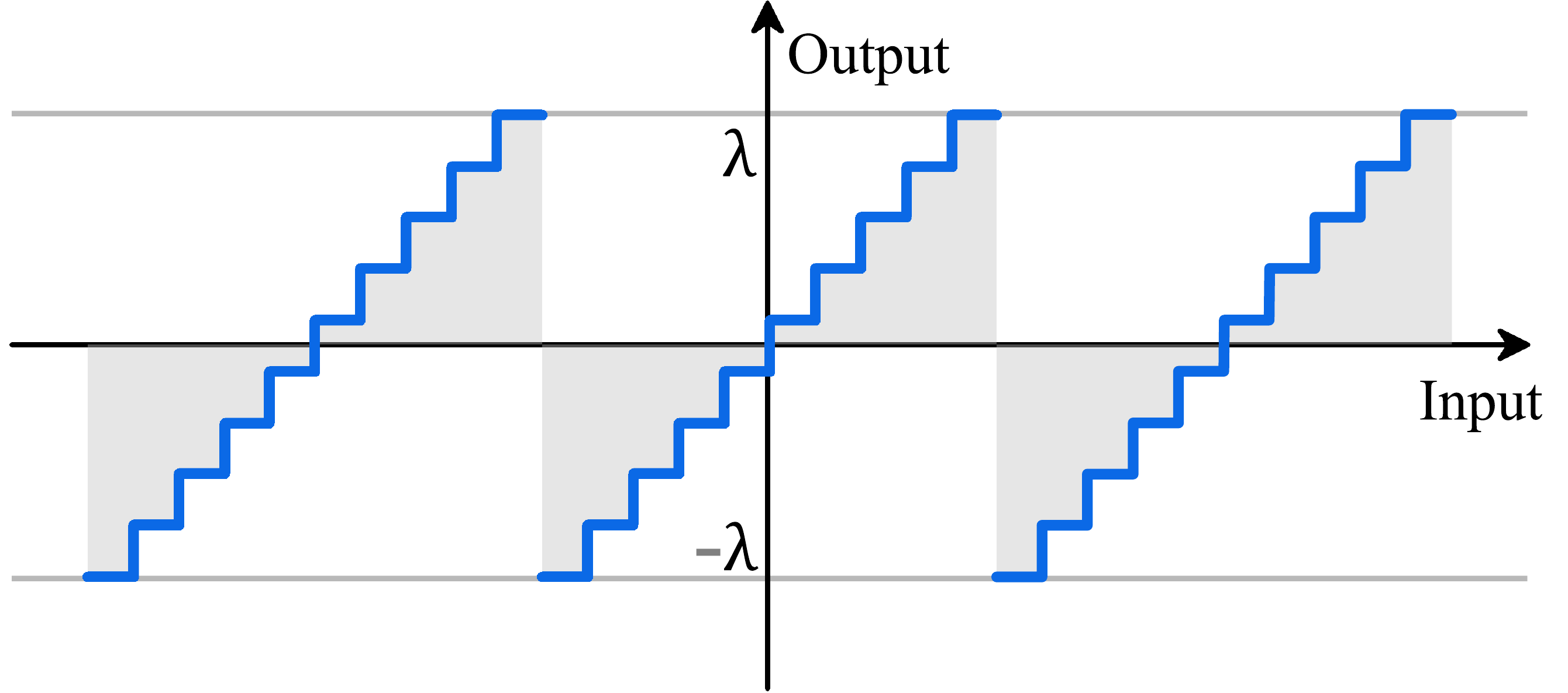}}
\vspace{-6pt}
\caption{Transfer functions of (a) modulo ADC and (b) conventional ADC.}
\vspace{-13pt}
\end{figure}

However, it is a highly ill-posed problem to recover an original signal from its modulo-reduced samples without knowing the residual information about folding locations. So far, most of researchers have been dedicating to solving this problem and proposing different recovery algorithms to guarantee successful recovery, such as N-order Difference (NoD) [1], Quotient Tracker (QT) and Ordinary Least Square (OLS) [2], and Wavelet Filtering with solving LASSO (WFL) [12]. In essence, these algorithms have similar principles. NoD, QT and OLS compute the difference values of adjacent modulo samples to identify the folding locations when the values are larger than a preset threshold. WLF algorithm identify modulo-folding locations by wavelet-filtering and solving a LASSO problem. 

Nevertheless, all the above algorithms are designed for the reconstruction from real-valued modulo measurements, none of them considers the impact of quantization, which is unrealistic because modulo samples must be quantized and mapped into a finite number of bits in ADC. Indeed, there are several literatures discussed about quantization effect. For instance, Bhandari \emph{et al}. [8] studied one-bit sigma-delta quantization in USF. This one bit represents the sign of the sample and a large number of samples are required for exact recovery, which is different from the general \emph{L}-bit scalar uniform quantizer that is commonly used in reality. In [6], Ordentlich \emph{et al}. gave a method to recover the input Gaussian signal from its quantized modulo version with the prior knowledge of the covariance matrix, which is, however, usually inaccessible in practical applications. Therefore, we focus on scalar uniform quantizers and make efforts on exact recovery without knowing the statistical information of input signals in advance.

\begin{figure*}
\centerline{\includegraphics[width=0.9\textwidth]{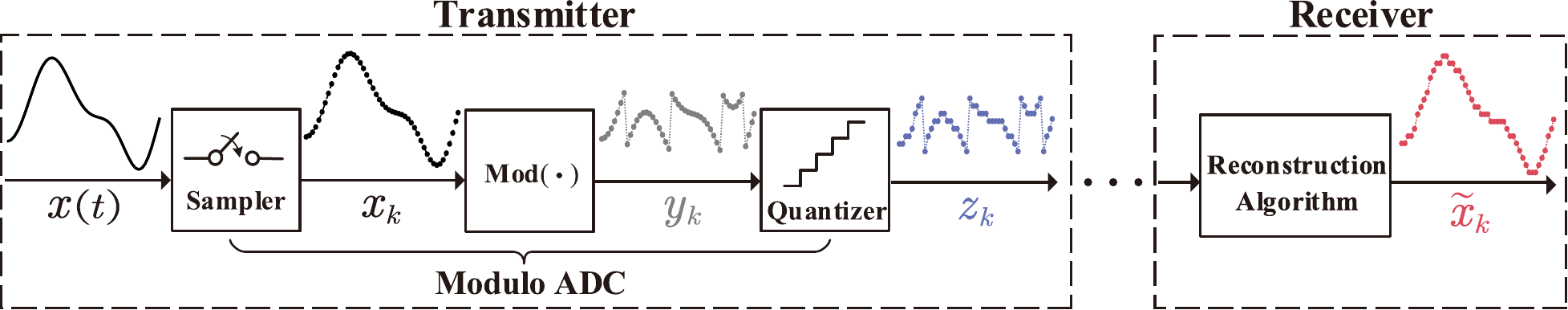}}
\vspace{-6pt}
\caption{The diagram of modulo ADC and reconstruction. At the transmitter, the original high-dynnamic-range signal $x\left(t\right)$ is sampled and modulo-folded into a low-dynamic-range modulo samples which are then be quantized and transmitted. At the receiver, the original signal can be reconstructed by a well-designed algorithm.}
\vspace{-13pt}
\end{figure*}

Intuitively, quantization results in an irreversible loss of information and using a very high precision quantizer is a straight way to reduce errors. But low-bit quantization may be preferred in data acquisition systems since it brings simplicity, robustness and cost efficiency. If quantization bit \emph{L} is low, severe error will occur inevitably and cause recovery failure, as demonstrated in Fig. 3(b). This failure leads to accumulation of errors and a rapid decline of signal-to-noise ratio of recovery signals, which motivates us to reexamine the sufficient conditions on sampling rate to guarantee exact recovery. The minimum sampling rate required in NoD algorithm is related to signal bandwidth $\Omega$ [1] while the minimum sampling rate in other algorithms are connected to the Lipschitz constants of original signals [2] [12]. Since the NoD algorithm offers an explicit trade-off when choosing sampling rate and computational complexity according to signal bandwidth, we choose the NoD algorithm in [1] for the following investigation.

To see this, our goal is to make the recovery algorithms robust to any quantized modulo samples so that ADC with any precision can be put into the implementation. We modeled the quantization noise in USF and explored the requirement on sampling rate to guarantee exact recovery in this model. We applied the proposed requirement to the procedure of sampling and recovered orginal signals using NoD. The results show that the recovery is stable and the errors induced by modulo-folding and recovery are negligible no matter how small the number of quantization bits is.

The rest of this paper is organized as follows. In Section II-A, a mathematical model for modulo ADC is introduced to characterize the impact of \emph{L}-bit uniform quantizer on USF. In Section II-B, we present a requirement on sampling rate to ensure successful recovery from quantized modulo samples. Section III shows the results of detailed statistical analysis to validate the proposed theory and outlines the trade-offs of the required minimum sampling rates, quantization bits and computational complexity. Conclusion is made in Section IV.
\vspace{-2.5pt}
\section{Mathematical Model}
\subsection{Modulo ADC}
{The diagram of the signal processing of a modulo ADC and reconstruction is demonstrated in Fig. 2. It includes taking the modulus of signal amplitude at the transmitter and reconstructing the input via a reconstruction algorithm at the receiver.

At the transmitter, given an $\Omega$-bandlimited analog input signal $x\left(t\right)$, sampled by a period of $T_s$ into a discrete-time signal $\left\{ x_k=x\left( kT_s \right) ,\mathrm{ }k\in \mathbb{Z} \right\}$ then mapped into a bounded area $\left[ -\lambda ,\mathrm{ }\lambda \right]$ for an appropriate threshold $\lambda\in\mathbb{R}^+$ by modulo-folding, we can obtain the modulo samples $y_k$, given by 
\vspace{-3pt}
\begin{equation}
y_k=\mathcal{M}_{\lambda}\left( x_k \right) =\left[ \left( x_k+\lambda \right) \mathrm{ mod }\;2\lambda \right] -\lambda
\label{eq.1}
\vspace{-3pt}
\end{equation}
where $\mathcal{M}_{\lambda}\left( \cdot \right)$ is centered modulo operation. The amplitude range after taking the modulo is $\left[ -\lambda ,\mathrm{ }\lambda \right]$, that is, the quantizer has the dynamic range of $\left[ -\lambda ,\mathrm{ }\lambda \right]$. 
The unknown residual should be specified as
\vspace{-3pt}
\begin{equation}
\zeta _k=x_k-y_k=\varepsilon _k\lambda ,\mathrm{ }\varepsilon _k\in \mathbb{Z}
\label{eq.2}
\vspace{-3pt}
\end{equation}
which is a piecewise-constant function and contains important information of folding locations. Although we do not record the residual $\zeta _k$, correct reconstruction of $\zeta _k$ is the core idea of recovery algorithms. Then the samples $y_k$ are quantized to $z_k$ such that each sample can be represented by a finite number of bits,
\vspace{-3pt}
\begin{equation}
z_k=\sigma \cdot \lfloor \frac{y_k}{\sigma} \rfloor +\frac{\sigma}{2}
\label{eq.3}
\vspace{-3pt}
\end{equation}
where $\sigma =\frac{\lambda}{2^{L-1}}$ is the step size of \emph{L}-bit scalar uniform quantization. At the receiver, the quantized modulo signal $z_k$ is received and $\tilde{x}_k$ is recovered from $z_k$ using a well-designed algorithm. The quantization error ${q}_k$, which is the major distortion of ADC, is expressed as
\vspace{-3pt}
\begin{equation}
q_k=z_k-y_k\in \left[ -\frac{\sigma}{2},\frac{\sigma}{2} \right] 
\label{eq.4}
\vspace{-3pt}
\end{equation}
Usually, $q$ is also regarded as quantization noise. For random signals like speech or images, $q$ is a random variable. 

Although it is not in the quantization process of ADC, the quantization of original signal $x_k$ should be specially noted as an implicit process. Combining (2) and (4), we can deduce the following equation,
\vspace{-3pt}
\begin{equation}
x_k=\zeta _k+y_k=\varepsilon _k\lambda +z_k-q_k
\label{eq.5}
\vspace{-3pt}
\end{equation}
For convenience, a quantization function $Q(\cdot)$ is defined as
\vspace{-3pt}
\begin{equation}
Q\left( x_k \right) =x_k+q_k=\varepsilon _k\lambda +z_k
\label{eq.6}
\vspace{-3pt}
\end{equation}
to denote the quantization version of the original signal $x_k$.}
\vspace{-12pt}
\subsection{Requirement for Exact Recovery from Quantized Signals}
{In this section, we will concretely study the impact of quantization on recovery condition. In [1], N-order difference operator is defined as $\Delta\! ^N$, where the first order difference is given by $\Delta y_k=y_{k+1}-y_k$. As shown in [1], in the absence of quantization, the principle for identifying folding locations is that if the following equation
\vspace{-3pt}
\begin{equation}
\mathcal{M}_{\lambda}\left( \Delta\!^Ny_k \right) =\Delta\!^Nx_k
\label{eq.7}
\vspace{-3pt}
\end{equation}
holds, the N-order difference values of the residual can be computed by combining (2) and (7) as $\mathcal{M}_{\lambda}\left( \Delta\!^Ny_k \right) -\Delta\!^Ny_k=\Delta\!^N\zeta _k$. Once we have $\Delta\!^N\zeta _k$, the residuals $\zeta _k$ can be obtained by repeated summation operation which is the inverse of repeated difference operation, then $x_k$ is obtained by (2).

In order to satisfy (7), the infinity norm of $\Delta\!^Nx_k$ must be less than $\lambda$. Hence, a sufficient condition was deduced by Taylor expansion and a series of inequalities scaling,
\vspace{-3pt}
\begin{equation}
\left\| \Delta\!^Nx_k \right\| _{\infty}\le \left( T_s\Omega \mathrm{e} \right) ^N\beta _x<\lambda 
\label{eq.8}
\vspace{-3pt}
\end{equation}
where $\beta _x=\left\| \mathrm{ }x\left( t \right) \right\| _{\infty}>\lambda $ is the upper bound of the infinity norm of original signal $x\left( t \right)$. The detailed derivation of (8) can be found in [1]. For simplicity, we denote the required minimum sampling rate of (8) as $f_{s1}$,
\vspace{-3pt}
\begin{equation}
f_{s1}=\frac{\Omega\mathrm{e}}{\sqrt[\uproot{5}\leftroot{-3}N]{\frac{\lambda}{\beta _x}}}
\label{eq.9}
\vspace{-3pt}
\end{equation}
It is worth noting that $f_{s1}$ must be higher than $\Omega \mathrm{e}$ because $\beta _x>\lambda $. 

However, we only have access to quantized modulo samples $z_k$ instead of unquantized $y_k$ at the receiver, the above requirement is no longer applicable. Note that (9) is derived from (7), now we should rewritten (7) as
\vspace{-3pt}
\begin{equation}
\mathcal{M}_{\lambda}\left( \Delta\!^Nz_k\mathrm{ } \right) =\Delta\!^NQ\left( x_k \right) 
\label{eq.10}
\vspace{-3pt}
\end{equation}
where $Q\left( x_k \right)$ is the quantization version of $x_k$. Given quantized modulo samples $z_k$, the above equation (10) holds only when the infinity norm of N-order differences of $Q\left( x_k \right)$ is less than $\lambda$, that is,
\vspace{-3pt}
\begin{equation}
\left\| \Delta\!^NQ\left( x_k \right) \right\| _{\infty}<\lambda  
\label{eq.11}
\vspace{-3pt}
\end{equation}
To ensure (11) holds, we firstly decompose $\Delta\!^NQ\left( x_k \right)$ into
\begin{equation}
\begin{aligned}
\Delta\!^NQ\left( x_k \right)&=\Delta\!^{N-1}\left( Q\left( x_{k+1} \right) -Q\left( x_k \right) \right) 
\\
&=\Delta\!^{N-2}\left( Q\left( x_{k+2} \right) -2Q\left( x_{k+1} \right) +Q\left( x_k \right) \right) 
\\
&=\cdots 
\\
&=\sum_{i=0}^N{\left( -1 \right) ^{N-i}C_{N}^{i}Q\left( x_{k+i} \right)}
\\
&=\sum_{i=0}^N{\left( -1 \right) ^{N-i}C_{N}^{i}\left( x_{k+i}+q_{k+i} \right)}
\\
&=\Delta\!^Nx_k+\sum_{i=0}^N{\left( -1 \right) ^{N-i}C_{N}^{i}q_{k+i}}
\label{eq.12}
\end{aligned}
\end{equation}
Next, the upper bound of $\Delta\!^NQ\left( x_k \right) $ is derived by
\vspace{-3pt}
\begin{equation}
\begin{aligned}
\left\| \Delta\!^NQ\left( x_k \right) \right\| _{\infty}&=\left\| \Delta\!^Nx_k+\sum_{i=0}^N{\left( -1 \right) ^{N-i}C_{N}^{i}q_{k+i}} \right\| _{\infty}
\\
&<\left\| \Delta\!^Nx_k \right\| _{\infty}+\left\| \sum_{i=0}^N{\left( -1 \right) ^{N-i}C_{N}^{i}q_{k+i}} \right\| _{\infty}
\\
&\le \left\| \Delta\!^Nx_k \right\| _{\infty}+\left\| \sum_{i=0}^N{C_{N}^{i}\cdot \frac{\sigma}{2}} \right\| _{\infty}
\\
&=\left\| \Delta\!^Nx_k \right\| _{\infty}+2^N\cdot \frac{\sigma}{2}
\\
&=\left\| \Delta\!^Nx_k \right\| _{\infty}+2^{N-1}\sigma 
\label{eq.13}
\end{aligned}
\end{equation}
We now bound $\left\| \Delta ^NQ\left( x_k \right) \right\| _{\infty}$ in magnitude by $\lambda$
\vspace{-3pt}
\begin{equation}
\left\| \Delta\!^NQ\left( x_k \right) \right\| _{\infty}<\left\| \Delta ^Nx_k \right\| _{\infty}+2^{N-1}\sigma \mathrm{ }<\lambda 
\label{eq.14}
\vspace{-3pt}
\end{equation}
to ensure (10) holds. According to (8) and (14), we have
\vspace{-3pt}
\begin{equation}
\left\| \Delta\!^Nx_k \right\| _{\infty}\le \left( T_s\Omega \mathrm{e} \right) ^N\beta _x<\lambda -2^{N-1}\sigma 
\label{eq.15}
\vspace{-3pt}
\end{equation}
Consequently, a new requirement for exact reconstruction is given as
\vspace{-3pt}
\begin{equation}
\left( T_s\Omega \mathrm{e} \right) ^N\beta _x<\lambda -2^{N-1}\sigma 
\label{eq.16}
\vspace{-3pt}
\end{equation}
In (16), the term $2^{N-1}\sigma$ is the errors induced by quantization and N-order difference operation. Similar to (9), we denote the minimum sampling rate of (16) as $f_{s2}$,
\vspace{-1pt}
\begin{equation}
f_{s2}= \frac{\Omega\mathrm{e}}{\sqrt[\uproot{5}\leftroot{-3}N]{\frac{ (\lambda -2^{N-1}\sigma)     }{\beta _x}}}
\label{eq.17}
\vspace{-1pt}
\end{equation}
Here, we obtain the new minimum sampling rate $f_{s2}$ required to successfully recover orignal signals from quantized modulo samples $z_k$.}

\vspace{-2.5pt}
\section{Numerical Results}
In this part, we carried out a series of numerical experiments to validate the required minimum sampling frequency $f_{s2}$. In order to evaluate the quality of the recovery signal, we firstly define a recovery signal-to-noise ratio (RSNR) in decibels as
\vspace{-3pt}
\begin{equation}
RSNR=10\log _{10}\!\:\left( \frac{\left\| x_k \right\| _{2}^{2}}{\left\| \tilde{x}_k-x_k \right\| _{2}^{2}} \right) 
\vspace{-3pt}
\end{equation}
where $x_k$ is the original signal and $\tilde{x}_k$ is the recovery signal, $\left\| \cdot \right\| _2$ denotes the $\ell _2$-norm. For comparison, we also define a signal-to-quantization noise ratio (SQNR) in decibels as
\vspace{-3pt}
\begin{equation}
SQNR=10\log _{10}\!\:\left( \frac{\left\| x_k \right\| _{2}^{2}}{\left\| q_k \right\| _{2}^{2}} \right) 
\end{equation}
where $q_k=Q\left( x_k \right) -x_k=z_k-y_k\in \left[ -\frac{\sigma}{2},\frac{\sigma}{2} \right]$. For a certain signal, the value of SQNR is determined by the number of quantization bits $L$, which is also the upper bound of RSNR. This means that if the recovery is successful, there is only quantization noise in the recovered signal.

To start with, we performed the recovery process from real-valued modulo samples with sampling rate setting as $f_{s1}$. The main parameters we used are $\beta _x=10$, $\lambda =2$, $N=1$. The results are shown in Fig. 3(a), the RSNR of recovery signal (red dots) obtained from the real-valued modulo samples (grey dotted line) is larger than 300 dB by using $f_{s1}$, which also means the reconstruction is perfect and the error induced by modulo-folding and recovery is negligible. 

\begin{figure}
		\centering
		\subfigure[]{\includegraphics[height=2.1cm,width=4.28cm]{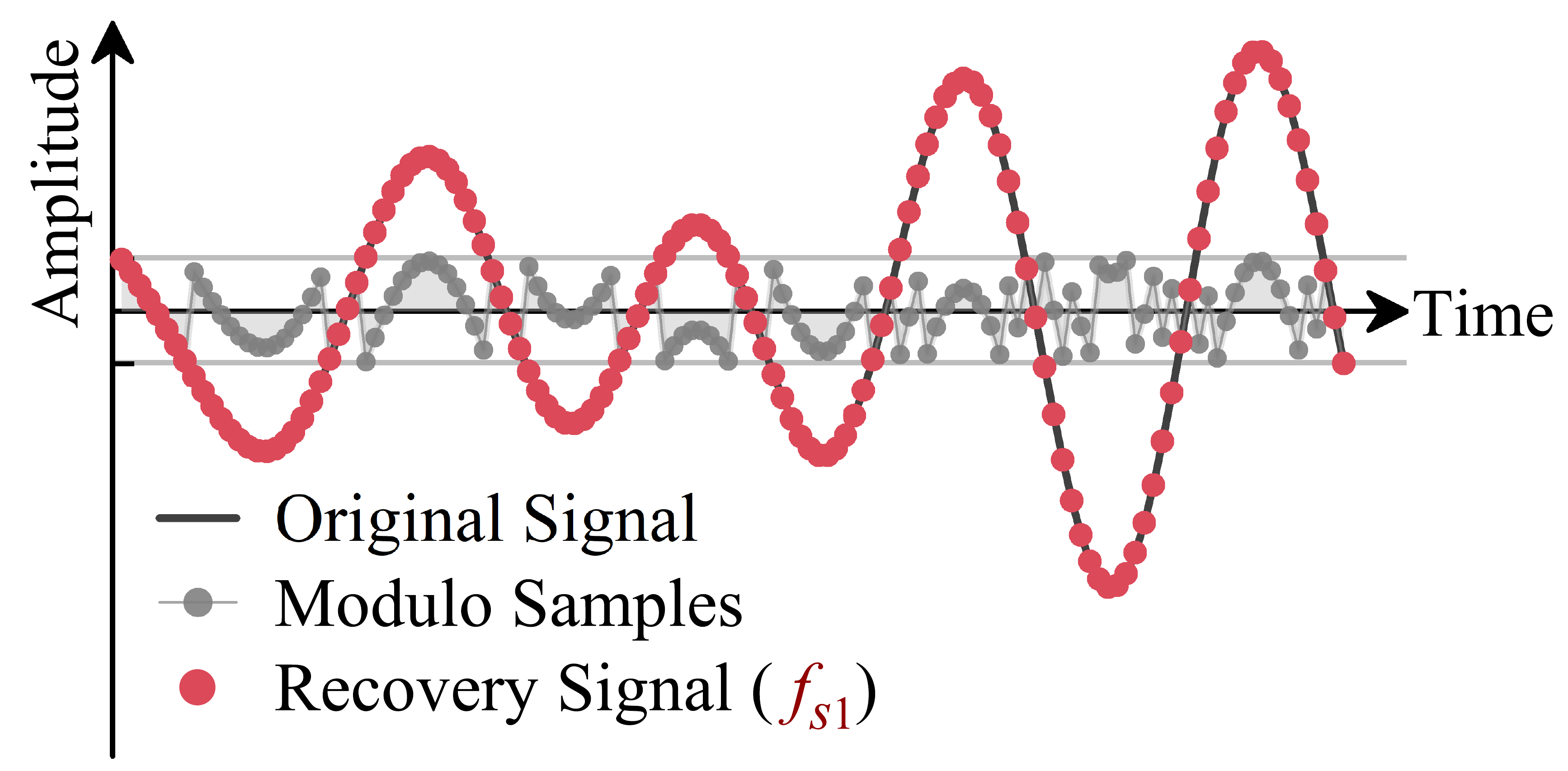}}
		\subfigure[]{\includegraphics[height=2.1cm,width=4.28cm]{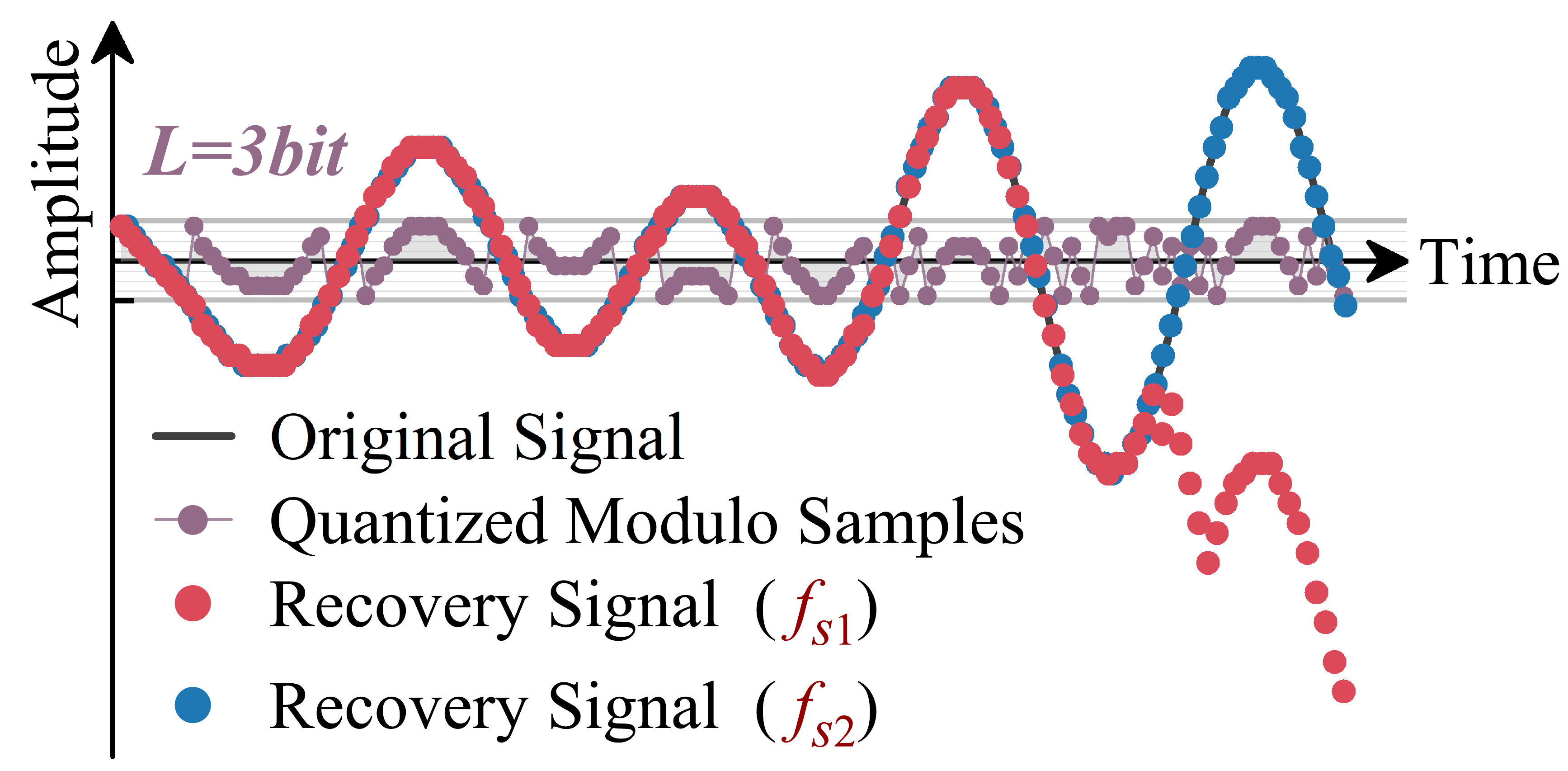}}
\vspace{-6pt}
\caption{(a) Perfect recovery from real-valued modulo samples. (b) Failure of recovery from 3-bit quantized modulo samples but succeeds after increasing sampling rate to $f_{s2}$.}
\vspace{-13pt}
\end{figure}

However, when modulo samples are quantized by a 3-bit ADC (violet dotted line), the recovery algorithm fails and the RSNR of recovery signal (red dots) is just 2.93 dB, as shown in Fig. 3(b). It is easily seen that the sampling rate $f_{s1}$ calculated from Eq. (9) is no longer applicable to quantized modulo samples in ADC systems. This recovery failure is mainly because the quantization changes the difference values of $N$ adjacent samples and leads to cumulative error.  

\begin{figure*}
\centerline{\includegraphics[width=\textwidth]{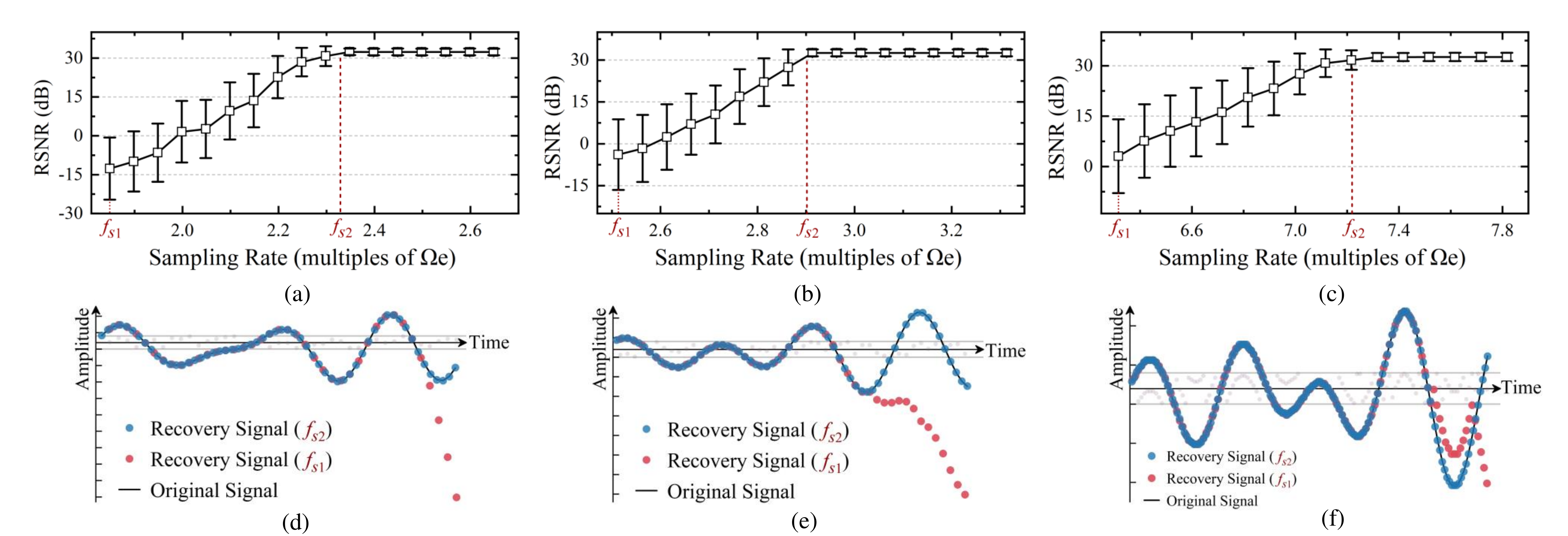}}
\vspace{-7pt}
\caption{The average RSNRs and standard deviation of RSNRs of 50 signals recovered by NoD where (a)$N=3$ (b)$N=2$ (c)$N=1$ at different sampling rates and the illustration of a randomly selected 4-bit quantized modulo signal sampled at $f_{s1}$ and $f_{s2}$ recovered by NoD where (d)$N=3$ (e)$N=2$ (f)$N=1$.}
\vspace{-15pt}
\end{figure*}

For the quantized modulo samples, we redid the recovery process with sampling rate setting as $f_{s2}$. It can be seen from the blue dots in Fig.3 (b) that the orignal signal is successfully restored with RSNR of 22.75 dB. Note that the SQNR for $L=3$ bit is about 23dB in this case, which the RSNR is very close to, indicating that the recovery is a success when modulo signal is sampled at the modified minimum sampling rate $f_{s2}$.

Next, in order to further investigate the applicability and robustness of the required minimum sampling rate $f_{s2}$, we randomly generated 50 input analog signals with different waveforms and bandwidths $\Omega$ but the same maximum amplitude $\beta _x$. For each original input signal, we sampled it at various sampling rate (from $f_{s1}$ to larger than $f_{s2}$) and took its modulus, then performed the recovery procedure and calculated the RSNR of the recovered signal $\tilde{x}$.

The results are shown in Fig. 4. The main parameters used in Fig. 4 are $\beta _x=12$, $\lambda =2$ and $\sigma =0.25$ ($L=4$ bit). The value of $N$ can theoretically take any positive integer as long as the right hand of inequality (12) is greater than zero. In this case, $N$ can take 1, 2 and 3. In Fig. 4(a)-(c), the squares and bars represent the average RSNRs and the standard deviation of RSNRs of 50 recovery signals respectively for different $N$. The horizontal axis indicates the sampling rate, with normalization to $\Omega\mathrm{e}$. The positions of minimum sampling rate $f_{s1}$ and the modified minimum sampling rate $f_{s2}$ are shown as well. As the sampling rate gradually increases from $f_{s1}$ to $f_{s2}$, the average RSNR increases from around 0dB to over 30dB and the length of error bar gradually decreases. When the sampling rate is greater than $f_{s2}$, the algorithm is stable and robust and the RSNRs are about 32dB that are equal to the SQNRs calculated from (19), which are higher than before because $L$ and $\beta _x$ is enhanced in this case. In Fig. 4(d)-(f), examples of recovery signals with sampling rate at $f_{s1}$ and $f_{s2}$ are selected from Fig. 4(a)-(c). It can be seen that when the sampling rate is $f_{s1}$, recovered signals (red dots) are distorted severely. When sampled at $f_{s2}$, all the modulo signals can be used to recover orignal signal perfectly, as shown in blue dots. In summary, the applicability and robustness of the sampling rate $f_{s2}$ have been validated.

Furthemore, we can see from Fig. 4 that $f_{s2}$ is different for different $N$ in NoD algorithm. The lower the $N$ (less complexity of computation), the higher the $f_{s2}$, which indicates that lower computational complexity can be traded off with more oversampling. The last investigation is to draw attention to the minimum sampling rates required with consideration of various quantization bits. Set the same parameters as before. The quantization bit $L$ ranges from 3 to 10 bits.The results in Fig. 5 provide further intuition on how $L$-bit quantization changes the required  minimum sampling rate $f_{s2}$ of each algorithm. When $L$ is low, the difference between $f_{s2}$ and $f_{s1}$ is relatively large. With the increase of quantization bits $L$, $f_{s2}$ gradually decreases and approaches to $f_{s1}$. In the case of low quantization bit, it is more necessary to carefully choose a suitable sampling rate. On the other hand, we can also observe that NoD algorithm indeed offers good tradeoff. For example, for any given $L$, $\lambda$ and $\beta _x$, the required minimum sampling rate is lower if $N$ is larger. However, larger $N$ adds more complexity in computation. It is also important to choose a suitable $N$ under different parameters setting. 
\vspace{-4pt}
\begin{figure}[H]
\centerline{\includegraphics[width=0.35\textwidth]{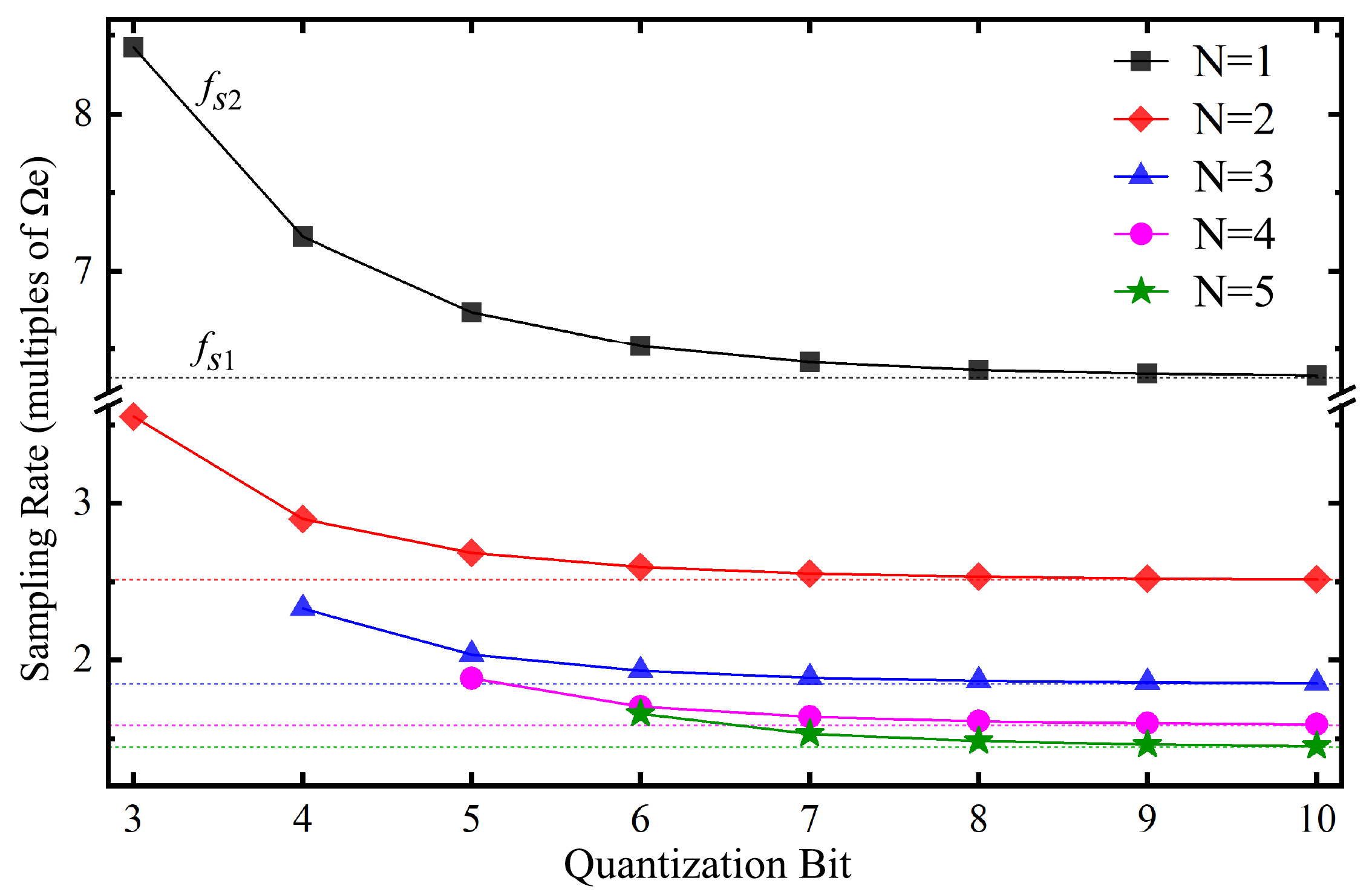}}
\vspace{-7pt}
\caption{The minimum sampling rate versus the number of quantization bits for different $N$. The solid lines correspond to $f_{s2}$ and the dashed lines correspond to $f_{s1}$.}
\vspace{-13pt}
\end{figure}
\section{Conclusion}
In this paper, we redefine the minimum sampling rate required for successfully recovering orignal signals from quantized modulo samples. The theory is verified by various numerical results that show the recovery is robust when sampling rate is above this lower bound. Furthermore, we give the relationship between sampling rates and quantization bits in various computational complexity. We hope these results serve as useful tools for practitioners when deciding which algorithms, quantization bits and sampling rates to use.

\end{document}